\documentstyle{ioplppt}                

\begin{document}

\title{Exact relativistic time evolution for a step potential barrier}
\author{Jorge Villavicencio\dag}

\begin{abstract}
We derive an exact analytic solution to a Klein-Gordon equation for a step
potential barrier with cutoff plane wave initial conditions, in order to
explore wave evolution in a classical forbidden region. We find that the
relativistic solution rapidly evanesces within a depth $2x_p$ inside the
potential, where $x_p$ is the penetration length of the stationary solution.
Beyond the characteristic distance $2x_p$, a Sommerfeld-type precursor
travels along the potential at the speed of light, $c$. However, no spatial
propagation of a main wavefront along the structure is observed. We also
find a non-causal time evolution of the wavefront peak. The effect is only
an apparent violation of Einstein causality.
\end{abstract}

\address{\dag\ Facultad de Ciencias, 
Universidad Aut\'{o}noma de Baja California \\ Apartado Postal 1880, 
Ensenada, Baja California, M\'{e}xico.}

%
%

\section{Introduction}

Since the early beginnings of quantum mechanics, the problem of particle
propagation in classical forbidden regions has been the subject of both
theoretical and experimental investigations. Over the years, several
non-relativistic approaches based in cutoff wave initial conditions{\em \ }%
have been introduced in the literature in order to investigate the
time-dependent features of wave evolution in evanescent media. Some of these
theoretical models \cite
{stevens83,morettipra92,morettipra92a,Buttiker,Muga&Butiker} were inspired
in the pioneering work of Sommerfeld and Brillouin \cite{Somm,Leoa}, while
others \cite{muga96,PRA97,PRA99,PRBRC} were based in the seminal work of
Moshinsky \cite{mm,mm1}, who a few decades ago started a fundamental
discussion on the non-relativistic and relativistic transient effects. These
models represent important steps towards the clarification of the dynamics
in classical forbidden regions, and a renewed motivation to explore this
problem has been recently stimulated by the issue of superluminal velocities
in photon \cite{chiao,chiaob} and microwave \cite{enders,nimtz} tunneling.
Hence, it is clear that a full relativistic approach to describe the wave
evolution in evanescent media is needed. Nevertheless, this has become a
complex problem due to the lack of exact analytical solutions to
relativistic wave equations with appropriate initial conditions. Among the
few works in the field \cite{Deutch,Buttiker,Muga&Butiker}, Deutch and Low 
\cite{Deutch} have provided a lucid description of barrier penetration of an
initial state given by a cutoff Gaussian wavepacket, based on a
one-dimensional Klein-Gordon equation. Although no exact relativistic
solutions were obtained, the issues of Einstein causality and superluminal
phenomena were rigorously discussed using approximate solutions.

In this paper we consider a model based on the Klein-Gordon equation, as in
the work of Deutch and Low \cite{Deutch}, for a potential step barrier and
cutoff plane wave initial conditions. We obtain an exact analytic solution
to the problem along the potential region and study the main features of
wave evolution, in particular the regime of transient effects at early times.

The paper is organized as follows. In section 2 we discuss the shutter
problem, and present the analytical derivation for the solution to a
relativistic wave equation for a step potential barrier. Section 3 deals
with a numerical example for the solution along the internal region of the
potential, and the results are discussed in section 4. Finally, in section 5
we present the summary and conclusions.

\section{The relativistic shutter problem}

To investigate the time evolution of cutoff plane wave in a classical
forbidden region, let us consider a classical field $\psi _r^s$ satisfying a
one-dimensional Klein-Gordon equation with a variable potential $V(x)$, as
in the model of Deutch and Low \cite{Deutch},

\begin{equation}
\frac{\partial ^2}{\partial x^2}\psi _r^s(x,k_r,t)=\frac 1{c^2}\frac{%
\partial ^2}{\partial t^2}\psi _r^s(x,k_r,t)+V_0(x)\psi _r^s(x,k_r,t).
\label{step}
\end{equation}

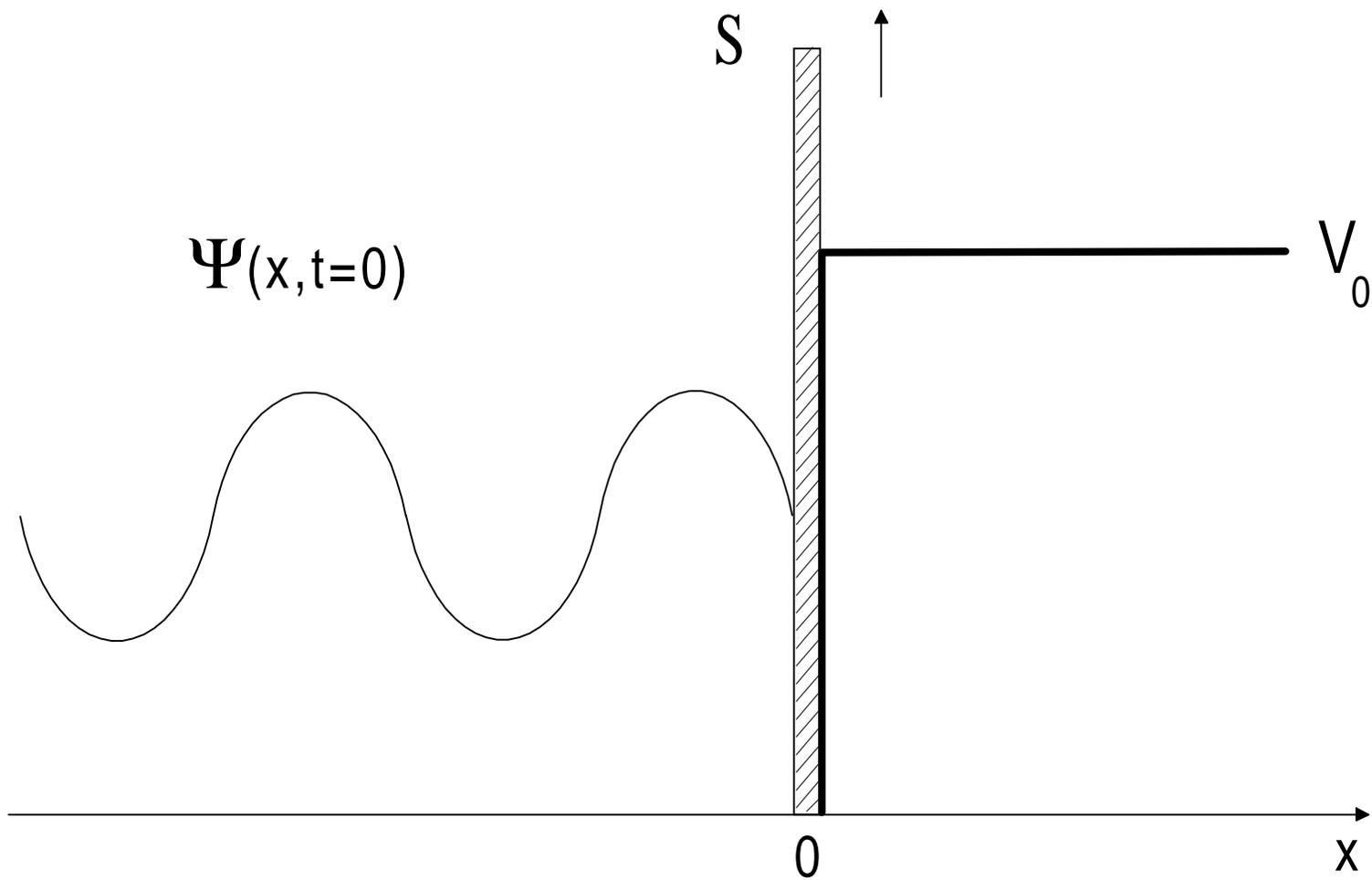
\begin{figure}[h]
\caption{Shutter problem for a potential step barrier $V_0.$ An initial
state $\psi (x,0)$ in the region $x<0$ is instantaneously released at $t=0$
by the removal of the shutter $S$.}
\label{fshutter}
\end{figure}

In our case, $V_0(x)$ is given by a step potential barrier,

\begin{equation}
V_0(x)=\left\{ 
\begin{array}{cc}
\mu _0^2, & x\geq 0, \\ 
0, & x<0,
\end{array}
\right.  \label{pot}
\end{equation}
where $\mu _0=(m_0c/\hbar )$, and the initial condition at $t=0$ corresponds
to a plane wave shutter \cite{PRA99} (see figure \ref{fshutter}), given by,

\begin{equation}
\psi _r(x,t=0)=\left\{ 
\begin{array}{cc}
e^{ik_rx}-e^{-ik_rx}, & x\leq 0, \\ 
0, & x>0.
\end{array}
\right.  \label{conref}
\end{equation}

The simplicity of our quasi-monochromatic initial state (\ref{conref}),
allows a closed analytical solution of the problem. It differs from that of
reference \cite{Deutch}, where a cutoff Gaussian wavepacket initial
condition was considered. Note that condition (\ref{conref}) comes from the
fact that for $t<0$, the solution for the left side of the shutter \cite
{note1} is given by $\psi _r(x,k_r,t)={\rm exp}[ik_r(x-ct)]-{\rm exp}%
[-ik_r(x+ct)]$, for $x<0$, and zero for $x>0$.

To obtain the solution for $x>0$ and $t>0$, we shall proceed along the same
lines as in our recent work \cite{PRA99}. We begin by Laplace transforming
the equation (\ref{step}) using the standard definition

\begin{equation}
\overline{\psi }(x,k_r,s)=\int\nolimits_0^\infty \psi (x,k_r,t)e^{-st}dt,
\label{laplace}
\end{equation}
with the initial condition given by equation (\ref{conref}). As a
consequence, one gets a pair of differential equations corresponding to the
regions $x>0$ and $x<0$. In order to obtain the transmitted wave function,
one must consider the matching conditions for the wave function and its
derivative at $x=0$. The Laplace transformed solution reads,

\begin{equation}
\overline{\psi }_r^s(x,s)=\frac{2E}{i\left( s+iE\right) \left( s+p\right) }%
e^{-px/c},  \label{steplap}
\end{equation}
where $p=(s^2+\mu _0^2c^2)^{1/2},$ and $E=k_r=(E_r/\hbar c)$ corresponds to
the relativistic energy $E_r$ given in reciprocal units of length.

The time dependent solution for $x>0$ is readily obtained by performing the
inverse Laplace transform of equation\ (\ref{steplap}) using the Bromwich
integral formula,

\begin{equation}
\psi _r^s(x,t)=\frac 1{2\pi i}\int\limits_{\gamma ^{\prime }-i\infty
}^{\gamma ^{\prime }+i\infty }\overline{\psi }_r^s(x,s)e^{st}ds,
\label{bromwich1}
\end{equation}
where the integration path is taken along a straight line $s=\gamma ^{\prime
}$ parallel to the imaginary axis in the complex $s-$plane. The real
parameter $\gamma ^{\prime }$ can be chosen arbitrarily as long as all
singularities remain to the left-hand side of $s=\gamma ^{\prime }$.

The integral (\ref{bromwich1}) expressed in this form, is difficult to
manipulate since the integrand (\ref{steplap}) has branch points at $s=\pm
i\mu _0c$. To surmount this difficulty, let us introduce the change of
variable, $-iu=(s+p)/\mu _0c$, which allows to eliminate the branch points.
Thus, $p=i\mu _0c(u^{-1}-u)/2$, and as a consequence, the Bromwich integral
may be written as,

\begin{equation}
\psi (x,t)=\frac 1{2\pi i}\int\limits_{i\gamma -\infty }^{i\gamma +\infty
}F(u)du,  \label{brom2}
\end{equation}
where the new integrand $F(u)$ is given by,

\begin{eqnarray}
F(u) &=&\frac{2E}{\mu _0}\frac{(1-u^2)}{u^2(u^2-2Eu/\mu _0+1)}  \nonumber \\
&&\times \text{exp}\{i\mu _0[u(x-ct)-u^{-1}(x+ct)]/2\}.  \label{integrand}
\end{eqnarray}

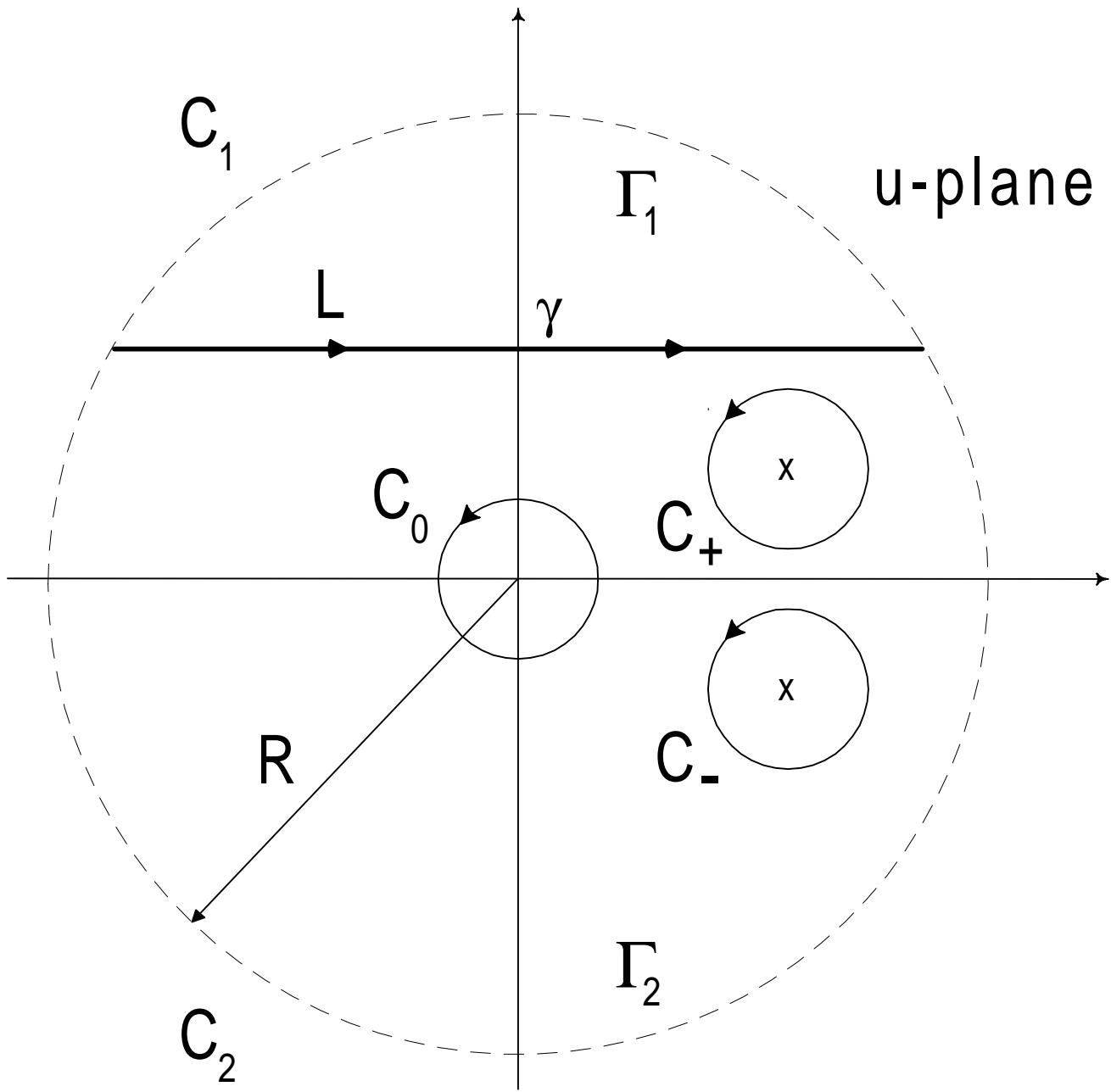
\begin{figure}[h]
\caption{ Integration contours $C_2=L+\Gamma _2+C_0+C_{+}+C_{-}$ and $%
C_1=L+\Gamma _1$, used to evaluate Eq. (7). The infinite
semicircles $\Gamma _1$ and $\Gamma _2$ (dashed line) correspond to the
cases $x>ct$ and $x<ct$, respectively. }
\label{fig2}
\end{figure}

Note that the branch points go into an essential singularity at $u=0$ and
two simple poles $u_{\pm }=(E\pm iq)/\mu _0$, where we defined $q=(\mu
_0{}^2-E^2)^{1/2}$. The integration in equation (\ref{brom2}) is performed
along a straight line $L$ parallel to the real axis cutting the positive
imaginary axis at $i\gamma $. We proceed to evaluate the above integral by
considering a closed Bromwich integration contour (see figure \ref{fig2}),
and Cauchy's residue theorem. For $x>ct$ we close the integration path $L$
from above, by a large semicircle $\Gamma _1$ of radius $R$, forming a
closed contour $C_1$. The contribution along $\Gamma _2$ vanishes as $%
R\rightarrow \infty $, and since there are no poles enclosed inside $C_1,$ $%
\psi (x,t)=0$ for $x>ct$. For the case $x<ct,$ we close the integration path
from below with a large semicircle $\Gamma _2$. The closed contour $C_2$
contains three small circles $C_{0,}$ $C_{+}$ and $C_{-}$ enclosing the
essential singularity at $u=0$ and the simple poles at $u_{+}$ and $u_{-},$
respectively. Hence by using Cauchy's theorem, it follows that,

\begin{equation}
\frac 1{2\pi i}\left[ \int\limits_{i\gamma -\infty }^{i\gamma +\infty
}-\int\limits_{\Gamma
_2}+\int\limits_{C_0}+\int\limits_{C_{+}}+\int\limits_{C_{-}}\right]
F(u)du=0.  \label{Cauchy}
\end{equation}

The integrals corresponding to the contours $C_{+}$ $C_{-}$ can be easily
evaluated, and yield the exponential contributions to (\ref{Cauchy}), namely,

\begin{equation}
\frac 1{2\pi i}\int\limits_{C_{\pm }}F(u)du=k_{\pm }e^{(\mp qx-iEct)},
\label{expon}
\end{equation}
where we defined $k_{\pm }=2E/(E\pm iq)$.

The contour integration for $C_0$ requires a more elaborate calculation,
since involves an essential singularity at $u=0$. For this case, we
introduce the change of variable given by $\omega =-iu\xi ^{-1}$, thus the
integral now becomes

\begin{equation}
\int\limits_{C_0}F(u)du=\int\limits_{C_0^{^{\prime }}}\frac{2E}{i\mu _0\xi ^3%
}\frac{(1+\omega ^2\xi ^2)exp[\eta (\omega -\omega ^{-1})/2]}{\omega
^2(\omega -\omega _{+})(\omega -\omega _{-})}d\omega ,  \label{C0}
\end{equation}
where $\omega _{\pm }=(E\pm iq)/i\mu _0\xi $. To carry out the integration,
first let us separate the integrand into partial fractions, and substitute
the well known formula for the Bessel generating function,

\begin{equation}
e^{\eta (\omega -\omega ^{-1})/2}=\sum_{n=0}^\infty \omega ^nJ_n(\eta
)+\sum_{n=1}^\infty (-1)^n\omega ^{-n}J_n(\eta )  \label{relation}
\end{equation}
and the series expansion,

\begin{equation}
(\omega _{\pm }-\omega )^{-1}=(\omega _{\pm })^{-1}\sum_{n=0}^\infty (\omega
/\omega _{\pm })^nJ_n(\eta ).  \label{relation2}
\end{equation}

The resulting integrals can be evaluated by means of the residue theorem.
For the case of an essential singularity, the residue may be determined by
computing explicitly the coefficient corresponding to $\omega ^{-1}$ from
the series expansion and their products. In that case, equation (\ref{C0})
becomes,

\begin{eqnarray}
\frac 1{2\pi i}\int\limits_{C_0}F(u)du &=&\left[ \frac{2iE}{\mu _0\xi }%
J_1(\eta )-k_{+}\sum_{n=0}^\infty (-1)^n\frac{J_n(\eta )}{(\omega _{+})^n}%
\right.  \nonumber \\
&&\left. -k_{-}\sum_{n=0}^\infty (-1)^n\frac{J_n(\eta )}{(\omega _{-})^n}%
\right] \text{.}  \label{C0_f}
\end{eqnarray}

Finally, substituting the results given by equations (\ref{C0_f}) and (\ref
{expon}) into equation (\ref{Cauchy}), the solution for the internal region
is,

\begin{equation}
\psi _r^s(x,t)=\left\{ 
\begin{array}{ll}
\psi _{+}(q)+\psi _{-}(q), & t>x/c \\ 
0, & t<x/c,
\end{array}
\right.  \label{stepfinal}
\end{equation}
with $\psi _{\pm }(q)$ defined as,

\begin{eqnarray}
\psi _{\pm }(q) &=&k_{\pm }\left[ e^{(\mp qx-iEct)}+\frac{iz_{\pm }}{2\xi }%
J_1(\eta )\right.   \nonumber \\
&&\left. -\sum\limits_{n=0}^\infty (\xi /iz_{\pm })^nJ_n(\eta )\right] .
\label{simplif}
\end{eqnarray}
In the above expression, $J_n(\eta )$ stands for the Bessel function of
order $n$. The other parameters are defined as: $\xi =\left[
(ct+x)/(ct-x)\right] ^{1/2},$ $\eta =\mu _0(c^2t^2-x^2)^{1/2}$ and $z_{\pm
}=(E\pm iq)/\mu _0$. From equation (\ref{stepfinal}) we see that the
solution obeys Einstein causality, i.e. no propagation faster than the speed
of light, $c$, is detected along the barrier region. In other words, an
observer located at an arbitrary position $x_0$ inside the barrier must wait
a time $t=(x_0/c)$ before detecting the arrival of the signal.

For the sake of completeness, let us now consider the asymptotic behavior of 
$\psi _r^s(x,t)$ for the cases $\mu _0\rightarrow 0$, $t\rightarrow \infty $
and $x\rightarrow ct$. From the solution we have just discussed, one may
recover the free propagation solution in the limit $\mu _0\rightarrow 0$.
This corresponds to let the variables $\eta \rightarrow 0,$ $q\rightarrow iE$%
. To illustrate the limit process in equation ( \ref{stepfinal}), let us
rewrite the solution by using equation (\ref{relation}), namely, 
\begin{eqnarray}
\psi _r^s(x,t) &=&k_{-}\left[ e^{(qx-iEct)}-J_0(\eta )\right.  \nonumber \\
&&\left. -\sum\limits_{n=1}^\infty (\xi /iz_{-})^nJ_n(\eta )\right] 
\nonumber \\
&&+k_{+}\left[ \sum\limits_{n=2}^\infty (z_{+}/i\xi )^nJ_n(\eta )\right] .
\label{as1}
\end{eqnarray}

As $\mu _0$ $\rightarrow 0$, the variable $J_0(\eta )\rightarrow 1$, and
since $(z_{-})^{-1}\rightarrow 0$ the first series on the right hand-side
clearly vanishes. It can be shown that the second series also vanishes, by
replacing the Bessel functions by their asymptotic values for small values
of the argument $\eta $,

\begin{equation}
J_n(\eta )\simeq 2^{-n}\eta ^n/n!.  \label{bessmall}
\end{equation}

Therefore, one obtains the solution for the free propagation case,

\begin{equation}
\psi _r^s(x,t)\rightarrow \left\{ 
\begin{array}{cc}
e^{ik_r(x-ct)}-1, & t>x/c, \\ 
0, & t<x/c.
\end{array}
\right.  \label{solibre}
\end{equation}

Note that the free case solution rises from zero only after a time $t=(x/c)$
fulfilling relativistic causality, and then oscillates periodically
thereafter.

In the case of the long-time limit $(t\rightarrow \infty )$, we have $\xi
\rightarrow 1$ and $\eta \rightarrow \infty $. From the asymptotic expansion
of $J_n(\eta )$ for large values of the argument $\eta $,

\begin{equation}
J_n(\eta )\simeq \frac 1{(\pi \eta /2)^{1/2}}cos[\eta -\frac 14(2n+1)\pi ],
\label{aslong}
\end{equation}
and therefore $J_n(\eta )\rightarrow 0$. One can see from equation (\ref
{simplif}) that the series in $\psi _{+}(q)$ vanishes. On the other hand, if
we rewrite $\psi _{-}(q)$ by means of equation (\ref{relation}), the
exponential term is canceled and the remaining series vanishes.
Consequently, $\psi _r^s(x,t)\ $goes into the stationary solution $\phi
_r^s(x,t)$ given by,

\begin{equation}
\phi _r^s(x,t)=k_{+}e^{-qx}e^{-iEct}.  \label{stationary}
\end{equation}

The asymptotic behavior near the relativistic cutoff, is obtained when $%
x\rightarrow ct$ in equation (\ref{stepfinal}). In this case we have $\eta
\rightarrow 0$, which allows us to substitute the asymptotic expansion (\ref
{bessmall}) in $\psi _{+}(q)$ (equation (\ref{simplif})). Thus, the series
of equation (\ref{simplif}) goes into an exponential function, which cancels
exactly with the exponential term, and as a result the solution $\psi _{+}(q)
$ goes like $iEJ_1(\eta )/\mu _0\xi $. From similar considerations on $\psi
_{-}(q)$, an identical expression is obtained and the approximate behavior
of $\psi _r^s(x,t)$ near the cutoff is given by,

\begin{equation}
\psi _r^s(x,t)\approx \frac{2iE}{\mu _0\xi }J_1(\eta ),  \label{bessel2}
\end{equation}
where for exactly the value $x=ct$, the above expression goes to zero
fulfilling relativistic causality.

\section{Examples}

In order to exemplify the evolution of the solution given by equation (\ref
{stepfinal}) along the evanescent region, we decided to study the properties
of $|\psi _r^s(x,t)|^2$ as a function of time $t$ and the position $x$. The
parameters for the system considered in all the cases for the present study
are: barrier height $\mu _0=1.542$ $nm^{-1}$, incidence energy $E_r=10.0$ $%
eV $ ( $E=5.064\times 10^{-2}$ $nm^{-1}$).

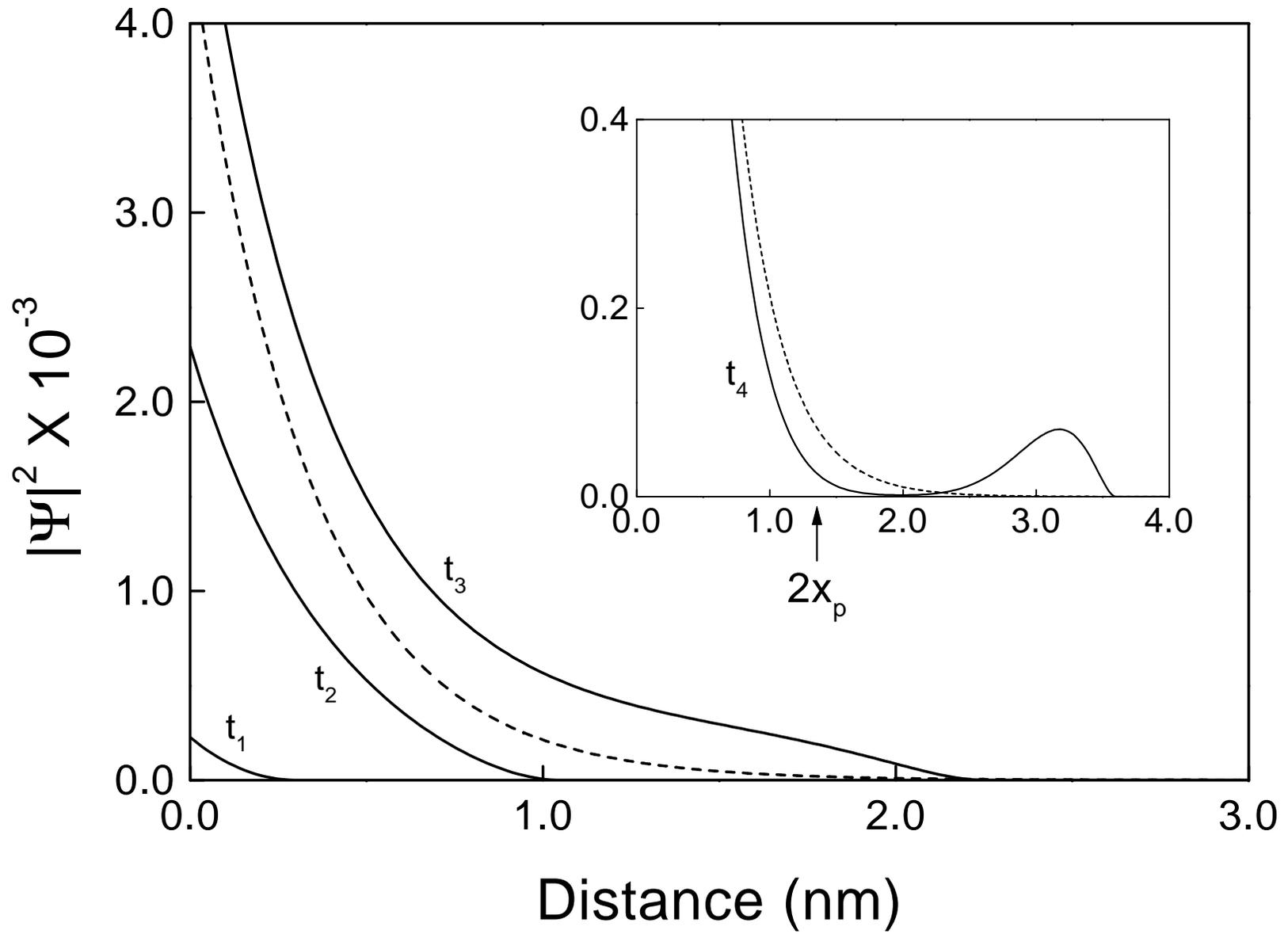
\begin{figure}[h]
\caption{ The birth of $|\psi _r^s(x,t)|^2$ (solid line) as
a function of distance $x$ for increasing values of time: $t_1=0.001$ $fs $, 
$t_2=0.0035$ $fs $ and $t_3=0.0075$ $fs $. Note that $|\psi _r^s(x,t)|^2$
fluctuates around the stationary solution $|\phi_r^s(x,t)|^2$ (dashed line).
The inset shows at a later time $t_4=0.012$ $fs $ the birth of a
Sommerfeld-type precursor near the relativistic cutoff at $x=3.0$ $nm$. }
\label{birth}
\end{figure}

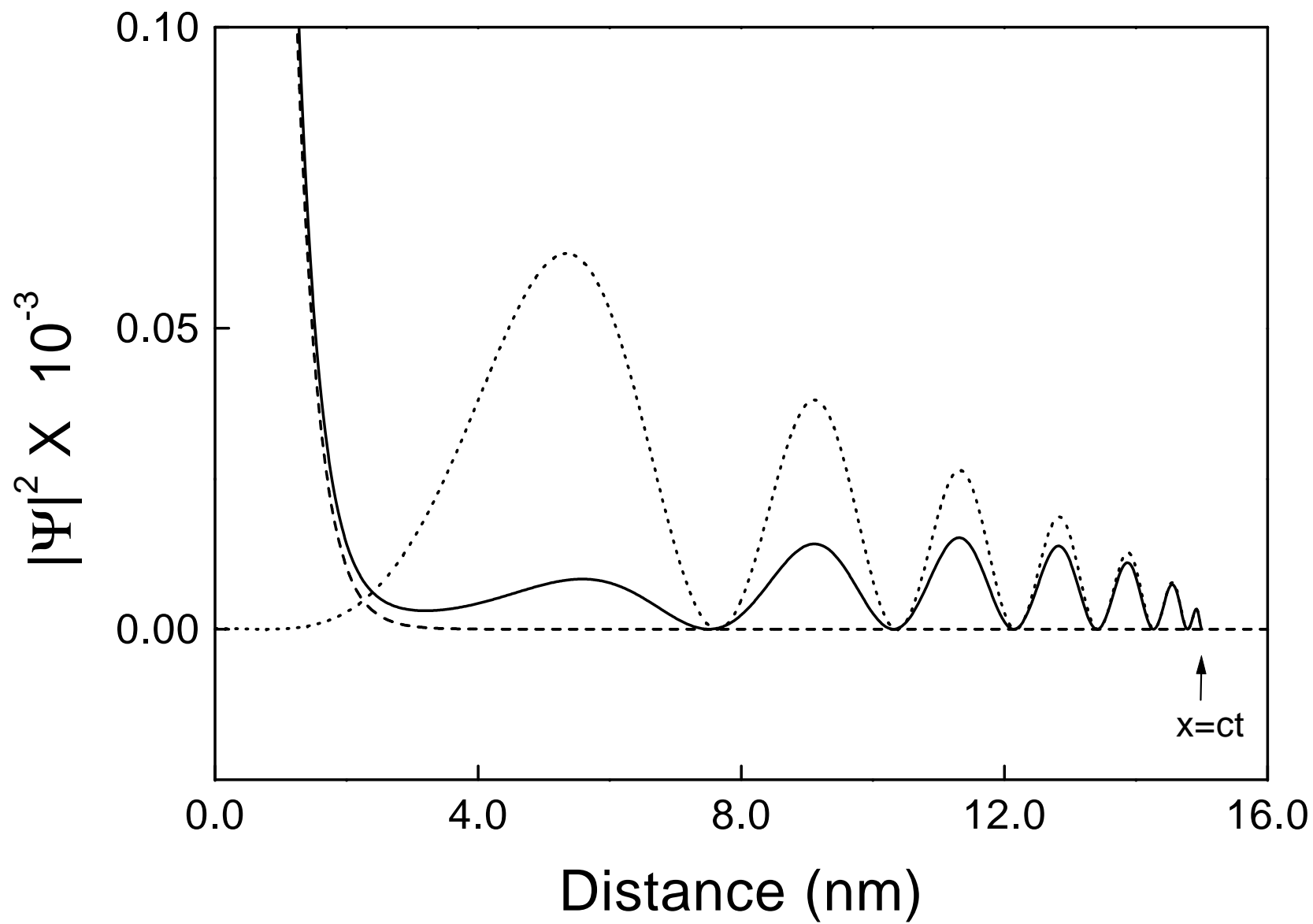
\begin{figure}[h]
\caption{ Plot of $|\psi _r^s(x,t)|^2$ (solid line) as a function of
distance $x$ for a fixed value of time $t=0.005$ $fs $. Notice that the wave
function exhibits a Sommerfeld-type precursor near the relativistic cutoff
at $x=15.0$ $nm$. The precursor is accurately described in the vicinity of
x=ct by a Bessel function (dotted line) given by equation (22). The
stationary solution $|\phi _r^s(x,t)|^2$ (dashed line) is also depicted in
the figure.}
\label{earlyt1}
\end{figure}

The first case corresponds to the spatial evolution of $|\psi _r^s(x,t)|^2$
along the dispersive region. In figure \ref{birth} we show at early times
the birth of the main wavefront as a function of the position, for
increasing values of time: $t_1=.001$ $fs$, $t_2=.0035$ $fs$ and $t_3=.0075$ 
$fs$. The solution rises as time goes on, and at $t_3$, $|\psi _r^s(x,t)|^2$
has already crossed over the stationary solution $|\phi _r^s(x,t)|^2$
(dashed line). The inset of figure \ref{birth} shows the crossover of $|\psi
_r^s(x,t)|^2$ at later time $t_4=.012$ $fs$. This behavior is relevant since
it indicates that the relativistic solution fluctuates around $|\phi
_r^s(x,t)|^2$, before reaching its asymptotic regime. At the inset, we can
also observe how the solution evanesces within a finite depth given
approximately by $2x_p=1.317$ $nm$, where $x_p=(1/q)$ is the {\it %
penetration length} of the stationary solution $|\phi
_r^s(x,t)|=|k_{+}|e^{-qx}$ (equation (\ref{stationary})). We find that
beyond $2x_p$ the solution exhibits a small maxima, corresponding to the
birth of a forerunner. In figure \ref{earlyt1} we depict the spatial
evolution of $|\psi _r^s(x,t)|^2$ (solid line) for a fixed value of $t=0.05$ 
$fs$. As can be seen in this example, the main part of the wave rapidly
evanesces in the potential region for small values of the position. However,
from approximately $2x_p$\ onwards, the solution exhibits an oscillatory
behavior before reaching the relativistic cutoff at $x=15.0$ $nm$,
corresponding to the earliest arrival of the signal at a point located
within the potential. The stationary solution $|\phi _r^s(x,t)|^2$ (dashed
line) is also included for comparison. It is interesting to note the
similarity of the oscillatory structure in figure \ref{earlyt1}, to the well
known Sommerfeld precursor \cite{Leob,precexpe}, which is one of the
essential features of wave propagation in dispersive media. Despite the fact
that Sommerfeld's approach is quite different from ours, the similarities go
beyond the numerical results. For instance, their asymptotic analysis showed
that the wave function is governed by a first order Bessel function near the
relativistic cutoff. Our analysis reproduces such behavior, which is given
by equation (\ref{bessel2}). For comparison, the value of the Bessel
function $J_1(\eta )$ modulated by the prefactor $2iE/\mu _0\xi $, is also
included in figure \ref{earlyt1} (dotted line). Note that if we define the
frequency of the oscillations of the precursor in terms of the distance
between successive zeros of $J_1(\eta )$, one sees from the definition of
the argument $\eta $\ that the value of the frequency depends only on the
position $x$\ and the potential $\mu _0$\ that characterizes the medium i.e.
the precursor frequency is independent of the incidence energy.

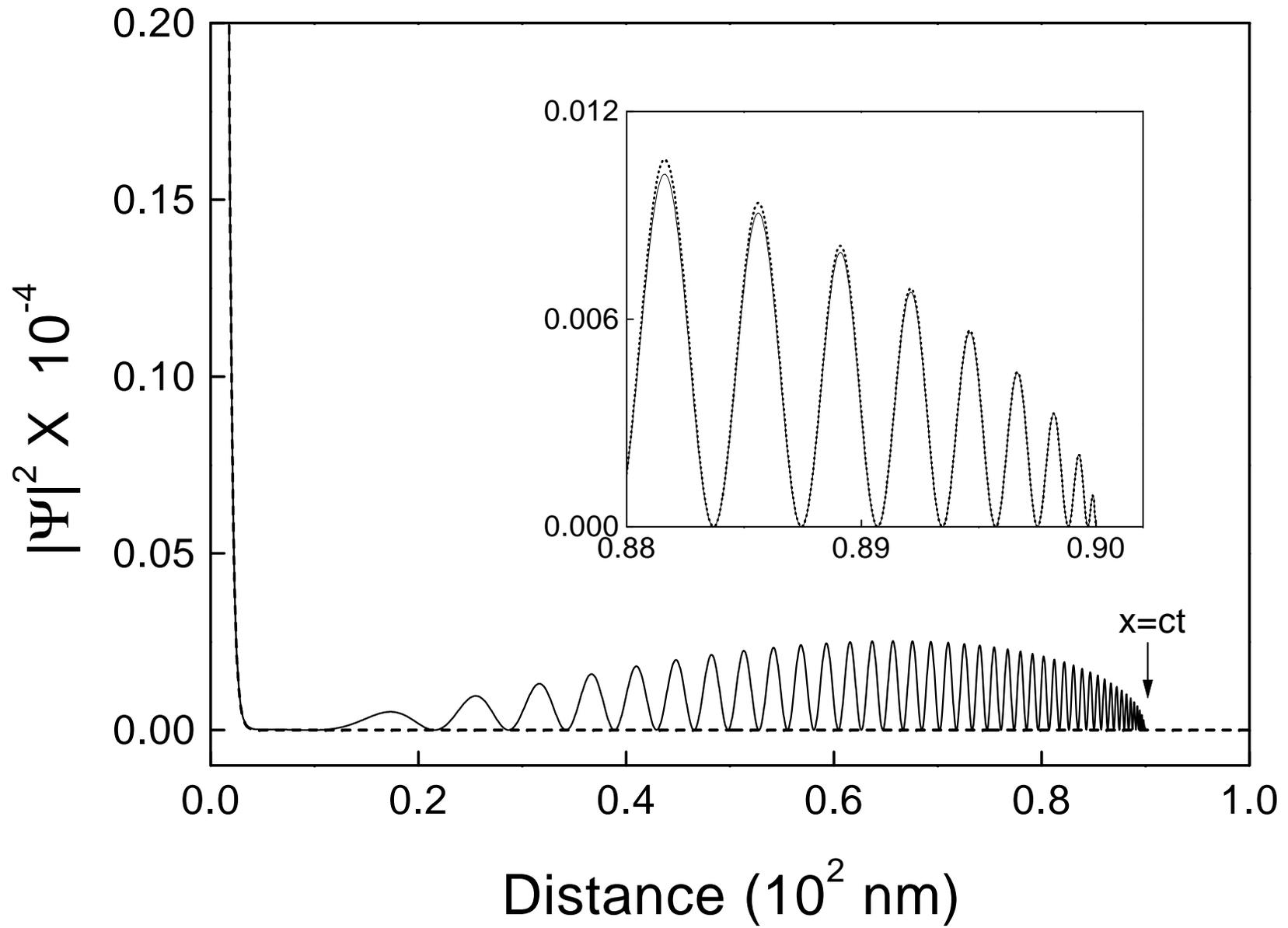
\begin{figure}[h]
\caption{ The main graph, as the previous one, shows the evolution of $|\psi
_r^s(x,t)|^2$ (solid line) at a later time $t=0.3$ $fs$. Note that the main
part of the wave reaches the stationary solution $|\phi _r^s(x,t)|^2$
(dashed line) at this short time. The small Sommerfeld-type precursor can be
observed near the relativistic cutoff at $x=90.0$ $nm$. At the inset we show
that the precursor (solid line), is well described by equation (22) (dotted
line). }
\label{earlyt2}
\end{figure}

In figure \ref{earlyt2} we show $|\psi _r^s(x,t)|^2$ (solid line) as a
function of the position $x$, at a subsequent time $t=0.3$ $fs$. We can see
that the solution reaches its stationary value $|\phi _r^s(x,t)|^2$ (dashed
line) for small values of $x$; nevertheless, near the relativistic cutoff at 
$x=90.0$ $nm$,\ the precursor exhibits a rich oscillatory structure. The
inset of figure \ref{earlyt2} illustrates the forerunner near the cutoff,
and shows that the asymptotic behavior is dictated by the Bessel function of
equation (\ref{bessel2}) (dotted line).

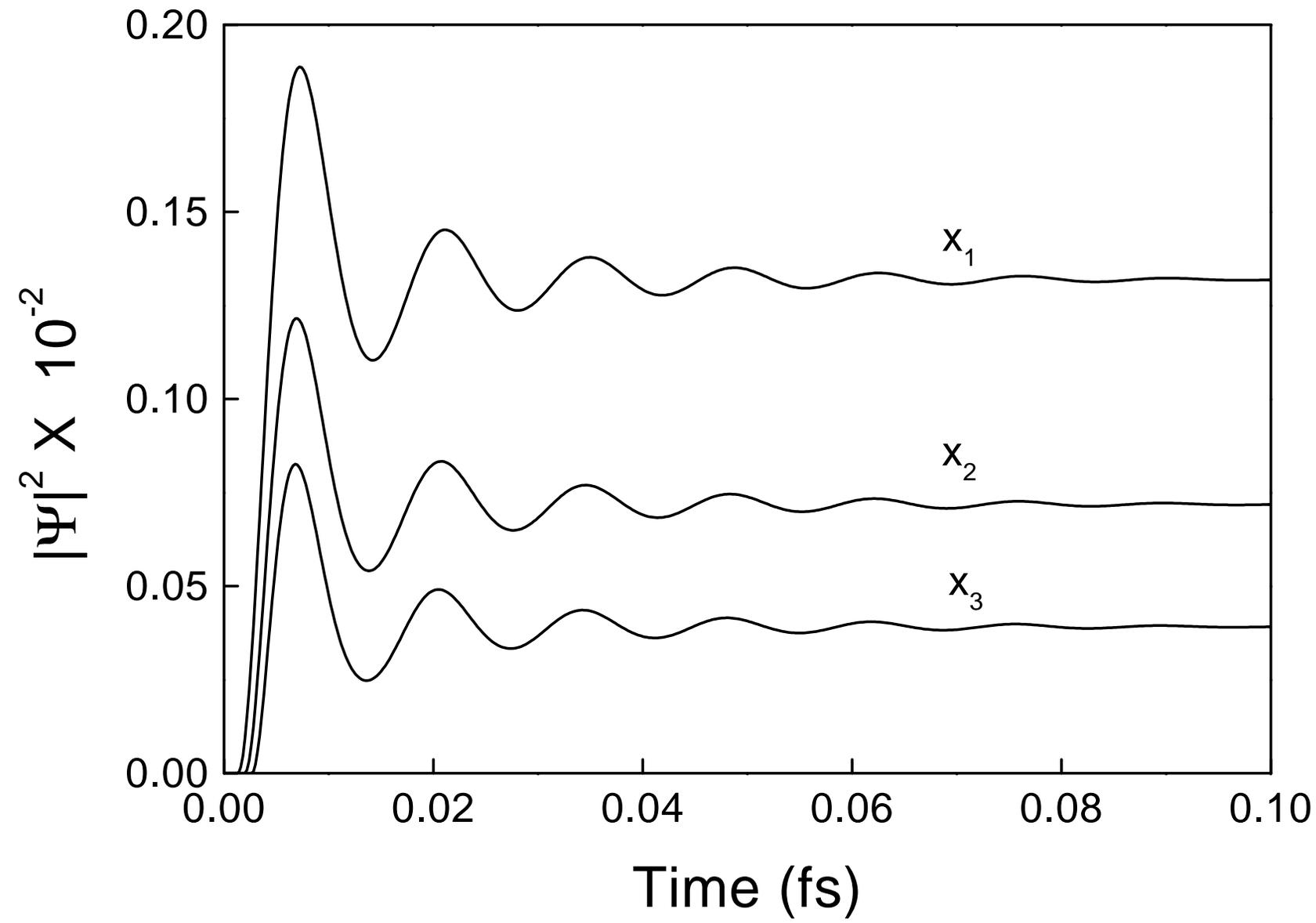
\begin{figure}[h]
\caption{ Time evolution of $|\psi _r^s(x,t)|^2$ for different values of the
position: $x_1=0.4$ $nm$, $x_2=0.6$ $nm$ and $x_3=0.8$ $nm$. Notice that the
transient behavior leading to the stationary regime exhibits an oscillating
pattern similar to the {\it diffraction in time} phenomenon.}
\label{diffraction}
\end{figure}

Up to here we have illustrated the spatial behavior of $|\psi _r^s(x,t)|^2$,
and some interesting features of the time evolution. In order to fully
explore the relevant features of the time evolution, in figure \ref
{diffraction} we plot $|\psi _r^s(x,t)|^2$ as a function of time at
different positions: $x_1=0.4$ $nm$\ , $x_2=0.6$ $nm$ and $x_3=0.8$ $nm$.
For all the curves depicted, as soon as $t>(x/c)$ the solution is different
from zero along the internal region, fulfilling relativistic causality. As
can be seen, the solution rises from zero at $t=(x/c)$ and grows
monotonically towards a maximum value, from which it starts to oscillate
thereafter, forming a pattern very similar to the diffraction in time
phenomenon \cite{mm}.{\bf \ }The concept of diffraction in time was
originally introduced by Moshinsky \cite{mm} while discussing the shutter
problem for the free particle Schr\"{o}dinger equation. He observed a
time-dependent oscillatory regime of the probability density near the
semiclassical wavefront that he named diffraction in time, in analogy to the
well known Fresnel optical diffraction. It is interesting to note the
resemblance of the oscillatory pattern in figure \ref{diffraction}, to the
diffraction in time phenomenon observed in the free propagation case \cite
{PRA99}. Moreover, in the low-energy regime ($\mu _0/E\gg 1$) the solution (%
\ref{stepfinal}) can be rewritten in a more concise form by using equation (%
\ref{relation}), namely, 
\begin{equation}
\Psi _r^s(x,t)\approx 2(E/V)\left[ U_3(i\eta /\xi ,\eta )-U_1(i\eta
/\xi,\eta )\right],  \label{flommel}
\end{equation}
where $U_1$\ and $U_3$\ are the Lommel functions of two variables \cite
{lommel2v}, widely used in connection with optical diffraction \cite
{bornwolf}. The resemblance to diffraction phenomena suggests that there
exists a more profound link; however, the physical implications of the
striking mathematical similarities found above deserves further study.

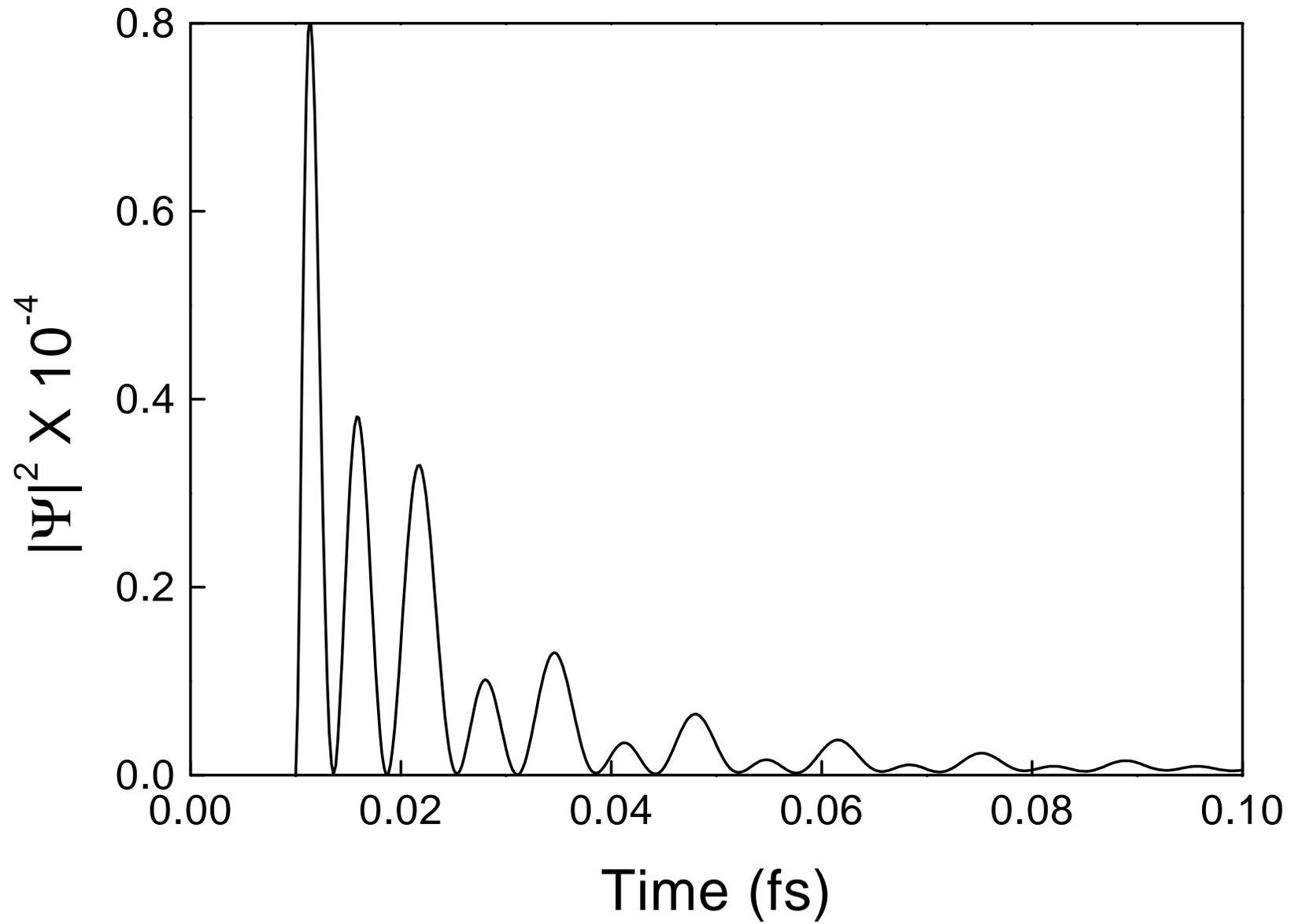
\begin{figure}[h]
\caption{ This graph illustrates $|\psi _r^s(x,t)|^2$ as a function of time
for a fixed value of the position $x_1=3.0$ $nm$, beyond $2x_p$. Notice that
in this case, the diffraction-type pattern clearly disappears. }
\label{nodiff}
\end{figure}

It is important to mention that the transient effect depicted in figure \ref
{diffraction} is observed in the low-energy regime i.e. $(\mu _0/E)\gg 1$;
this condition is satisfied in the present example, where the effect was
observed for values of the ratio{\bf \ }$(\mu _0/E)\simeq 30${\bf .}
Moreover, we only observed the phenomenon in the regime of small values of
the position $x$ where the solution decays in the potential region i.e. $%
x<2x_p$. From values greater than $x\simeq 2x_p$ the solution enters into a
different oscillatory regime, and the diffraction-type pattern begins to
disappear. In figure \ref{nodiff} we illustrate the inhibition of the
diffraction-type pattern for a fixed value of the position $x=3.0$ $nm$.
Clearly, it fades out and is replaced by a series of oscillations, which
register the fast crossing of the precursor at $x=3.0$ $nm$,\ and the
remaining forerunners.

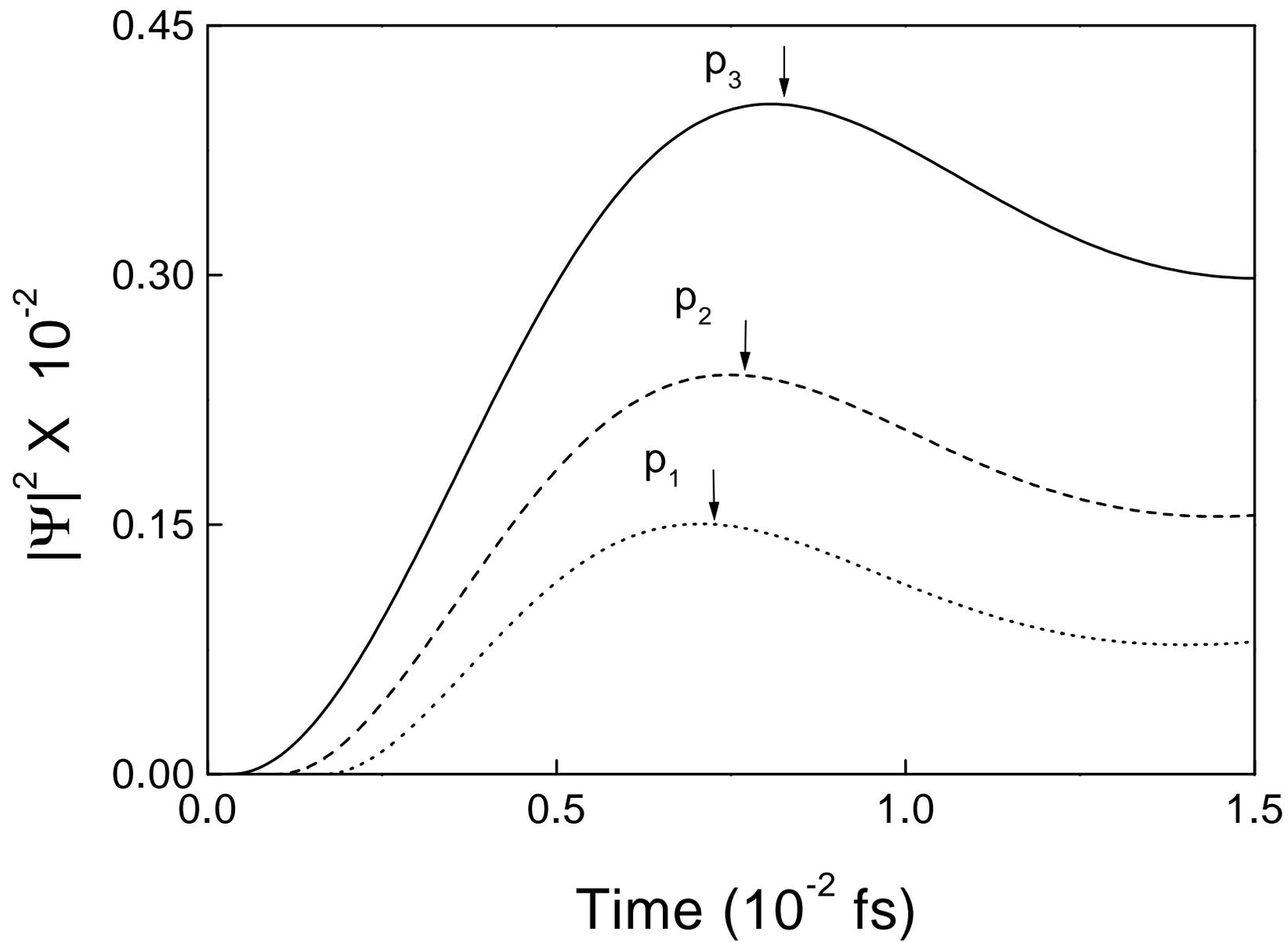
\begin{figure}[h]
\caption{ The time evolution of $|\psi _r^s(x,t)|^2$ in order to
exhibit the main peak shift of the wave, for different values of the
position: $x=0.1$ $nm$ (solid line), $x=0.3$ $nm$ (dashed line) and $x=0.5$ $%
nm$ (dotted line). The corresponding peak positions are $p_3$, $p_2$ and $%
p_1 $, respectively. Despite the fact that the three curves fulfill
relativistic causality, the wave front main peak exhibits an apparent
violation of Einstein causality. }
\label{peakshift}
\end{figure}

There is another interesting feature in the time evolution of $|\psi
_r^s(x,t)|^2$ that can be appreciated in figure \ref{peakshift}, in which we
plot $|\psi _r^s(x,t)|^2$ as a function of time in the main peak region.
Surprisingly, the maximum peak of the wave appears on $x=0.5$ nm\ (dotted
line) earlier than the peak at $x=0.3$ nm\ (dashed line) and $x=0.1$ nm\
(solid line). This relative {\it time shift} of the wave peak is an apparent
violation of relativistic causality, and can be interpreted as a non-causal
behavior. This comes from the fact that we are comparing the maximum wave
peak at different positions. However, we observe that the wavefront always
fulfills Einstein causality, and no signal travels faster than $c$ in the
dispersive region. Therefore, the observed shift of the main peak may be
interpreted as a reshaping of the wave and not as a genuine violation of
relativity.

It is interesting to mention that we have observed a similar non-causal
behavior in the probability density along a classical forbidden region of a
rectangular potential barrier, within a non-relativistic framework.
Moreover, some authors have also reported non-causal phenomena in
electromagnetic evanescent modes \cite{nimtz}.

\section{Discussion}

The possibility of describing the wave evolution from the transient to the
stationary regime, offers a clear advantage over the asymptotic methods of
solution available in the literature, for which the short and intermediate
transient regimes are inaccessible. In what follows, we shall discuss the
new features in the dynamical process of evanescent waves observed in the
previous section.

The buildup of $|\psi _r^s|^2$ exhibits a very interesting behavior; the
solution instead of just grow monotonically towards the stationary solution $%
|\phi _r^s|^2$, fluctuates around such value before reaching the asymptotic
regime, as it is shown by the series of curves of $|\psi _r^s|^2$ versus $x$
at different times (see figure \ref{birth}). The effect of these
fluctuations in a plot of $|\psi _r^s|^2$ versus $t$ ($x$ fixed) is
manifested as a series of oscillations similar to a diffraction in time
pattern, see figure \ref{diffraction}. The inset of figure \ref{birth} shows
that beyond a certain distance, an interesting structure of the wave
appears; this is the birth of the Sommerfeld-type forerunner which travels
along the potential region, as illustrated in figures \ref{earlyt1} and \ref
{earlyt2}; the head of this signal propagates at the speed of light and can
reproduced by the first order Bessel function $J_1(\eta )$ (dotted line).
The birth of the forerunner is an important event since its propagation at
longer times becomes the dominant process; this is also the case in the
context of different relativistic and non-relativistic approaches \cite
{Buttiker,Muga&Butiker}, where the characterization of the forerunners has
been recently emphasized.

At early times and for small values of the position, the main front of the
wave decays exponentially along the potential. As time goes on, the dynamics
is dominated by the propagation of the forerunners since the main front
rapidly reaches its asymptotic value without propagation. Thus, one may
speak of two regimes, which as discussed in the previous section, are
characterized by $2x_p$\ where $x_p$\ is the penetration length. If we
choose a position $x>2x_p$ and wait for the main wavefront, instead of
detecting its arrival we would only register the fast crossing of the
precursor (see figure \ref{nodiff}). The absence of main wavefront
propagation in the evanescent region is in agreement with a series of works 
\cite{muga96,Ranfagni,Jauhjo,Teranishi}, which have questioned the existence
of semiclassical wavefront propagation proposed by Stevens \cite{stevens83}
and supported by Moretti \cite{morettipra92,morettipra92a}.

Another important result of this work is the non-causal peakshift exhibited
in figure \ref{peakshift}. The non-causal behavior observed here is a
consequence of the reshaping of the wave; reshaping effects have also been
observed in the context of wavepacket evolution within both relativistic 
\cite{Deutch} and non-relativistic \cite{Krenslin} approaches. The role of
this effects and the issue of non-causal behavior has been recently
discussed by Deutch and Low \cite{Deutch} for the case of Gaussian
wavepacket evolution in the transmitted region of a potential barrier, based
on approximate solutions to the Klein-Gordon equation. Although the barrier
and the step potential are different systems, both exhibit an evanescent
region; hence, we believe that the non-causal behavior observed in reference 
\cite{Deutch} could be related to a reshaping process occurring inside the
barrier, similar to the reshaping observed in the step discussed in our
model. However, in order to investigate such a reshaping inside the barrier,
the solution of the Klein-Gordon equation is required for the internal
region. In this respect, the analytical techniques used in this work may
provide a suitable method of solution to tackle this fundamental problem;
nevertheless, this not an easy task since the extension of our model to the
case of a barrier of finite width involves more complicated analytical
properties of the solution due to the presence of resonances.

\section{Summary and conclusions}

We have derived an exact analytical solution to a Klein-Gordon equation for
a step potential barrier, using a cutoff plane wave initial condition. To
our knowledge, this is the first model which allows a closed solution for
the description of relativistic transient effects in a classical forbidden
region.

The main features of the spatial and time evolution along the evanescent
region can be summarized in the following points: (i) We found a regime
where the solution is exponentially suppressed and thus, decays as a
function of $x$ along the potential. This main part of the wave does not
propagate along the structure. The regime is characterized by a region
extending from $x=0$ to $x\approx 2x_p$, where $x_p=(1/q)$ is the
penetration length of the stationary solution (\ref{stationary}). However,
from $x\approx 2x_p$ onwards, the solution exhibits an oscillating pattern
near the relativistic cutoff, traveling at the speed of light, $c$ which can
be identified as a Sommerfeld-type precursor. Also, within the finite depth $%
2x_p$, we found in the low-energy situation that the time evolution of $%
|\psi _r^s(x,t)|^2$, exhibits a transient effect similar to the diffraction
in time phenomenon \cite{mm}. (ii) We showed that along the internal region,
there exists a time shift associated to the main peak of the wave function,
that can be interpreted as a non-causal behavior along the classical
forbidden region. This of course is only an apparent violation of
relativistic causality, since in our model the wavefront satisfy always
Einstein causality i. e. no signal travels faster than the speed of light.

The relevance of these results, comes from the fact that our findings may be
of interest to elucidate on the problem of wave propagation in finite width
potentials.

\ack

I would like to thank Gast\'{o}n Garcia-Calder\'{o}n and Alberto Rubio, who
suggested the shutter model approach to solve this problem. I am indebted to
them for encouragement and valuable discussions. I would like to thank also
Marcos Moshinsky for fruitful discussions on the shutter problem. Useful
discussions with Roberto Romo are gratefully acknowledged. This work was
supported financially by Conacyt-M\'{e}xico under Contract No.
431100-5-32082E.

\Figures

\begin{thebibliography}{99}
\bibitem{note1}  Note that the shutter is a device that aids to visualize
the initial condition and hence it is not part of the system.
\end{thebibliography}
\end{document}